# LIMIT THEOREMS FOR DECOHERENT TWO DIMENSIONAL QUANTUM WALKS

**CLEMENT AMPADU**

31 Carrolton Road
Boston, Massachusetts, 02132
U.S.A.
e-mail: drampadu@hotmail.com

**Abstract**

In this paper we consider the model with decoherence operators introduced by [Brun,T.A, et.al, Phys.Rev.A 67 (2003) 032304] which has recently been considered in the two-dimensional setting by [Ampadu,C., Brun-Type Formalism for Decoherence in Two Dimensional Quantum Walks, Communication in Theoretical Physics To Appear, arXiv:1104.2061 (2011)] to obtain the limit of the decoherent quantum walk.

**KEY WORDS:** limit theorems, decoherence, quantum random walk

**PACS:** 05.30.-d, 03.67.Lx, 05.40.-a

## I. Introduction

As is well known the physical implementation of the quantum walk faces many obstacles including environmental noise and imperfections collectively known as decoherence. Apart from the review on the decoherent quantum walk in [1], other studies on the decoherent quantum walk can be found in[2-26] and have been reviewed by the author of the present paper in [27]. In [28], the decoherent quantum random walk on the 1-dimensional integer lattice $Z$ is studied, leading to expressions for the first and second moments of the position distribution, it is also shown in the long time limit that the variance grows linearly with time with the diffusive character. In [27] the Brun type decoherence is extended to the two dimensional setting providing generalizations with wide range of applications. The generalized first and second moments for the decoherent quantum walk is obtained, the Brun formalism for the quantum walk is also treated. In the presence of broken line noise, the diffusive character of the walk is studied. It is conjectured that the diffusion coefficient in the quantum realm varies directly as $1-p$, and inversely as $p^2$, where $p$ is the probability of adjacent broken link at a given site in the walk. The conjecture if holds true implies the diffusion coefficient of the decoherent quantum walk is always larger than the diffusion coefficient in the classical case.

As the author of the present paper pointed out in [27], due to the complexity of the calculations, the pure analytic papers on the decoherent quantum walk have been given little attention in the literature. Moreover in [29] it is found that the complicated form of the superoperator in [28] makes it difficult to obtain the limit of the decoherent quantum walk. However, this difficulty is overcome by analyzing the characteristic function of the position probability distribution.

In this paper we follow the convention of obtaining the limit of the decoherent quantum walk by analyzing the characteristic function. This paper is organized as follows. In Section II some basic notions about the decoherent quantum walk is presented. Let $R(q,t) \equiv \left\langle e^{iq(x+y)} \right\rangle_t = \sum_{x,y} e^{iq(x+y)} P(x,y,t)$ be the characteristic function of the position probability distribution $P(x,y,t)$. In Section III we present the main result about $P_t(x,y,t) = P(\sqrt{t}x, \sqrt{t}y, t)$, $x, y \in \frac{Z}{\sqrt{t}}$, the rescaled probability mass function. We show if 1 is an eigenvalue of the superoperator with multiplicity 1, and there is no other eigenvalue whose modulus equals 1, then the characteristic function converges to a convex combination of normal distributions. In particular we show $P_t(x,y,t)$ converges in distribution to a continuous convex combination of normal distributions. Section IV is devoted to the conclusions, there an interesting problem is proposed which in a sense concerns illustrating the results of Section III.

## II. Definitions

Consider the quantum random walk on the general square lattice $Z^2$. Let the state space be given by $H_P \otimes H_C$, where $H_P$ denotes the position space and $H_C$ denotes the coin space. Let the basis for the position space be given by $\{|x,y\rangle : x,y \in Z\}$, and let the basis for the coin space be given by $\{|L\rangle, |R\rangle, |U\rangle, |D\rangle\}$, where $L,R,U,D$ represent the left, right, upward, and downward chirality states respectively. Let the shift operators in $H_P$ be defined as follows: $R^+|x,y\rangle = |x+1,y\rangle$, $L^-|x,y\rangle = |x-1,y\rangle, U^+|x,y\rangle = |x,y+1\rangle$, and $D^-|x,y\rangle = |x,y-1\rangle$, where $R^+, L^-, U^+, D^-$ are unitary shift operators on the particle position. Let $P_R, P_L, P_U, P_D$ be the orthogonal projections on the coin space $H_C$ spanned by $\{|L\rangle, |R\rangle, |U\rangle, |D\rangle\}$, where $P_R + P_L + P_U + P_D = I$. Let $M$ be the unitary transformation on $H_C$, then the evolution operator of the quantum walk is given by $G = \{(R^+ \otimes P_R) + (L^- \otimes P_L) + (U^+ \otimes P_U) + (D^- \otimes P_D)\}(I \otimes M)$. The eigenvectors $|k_x,k_y\rangle$ of $R^+, L^-, U^+, D^-$ are given by $|k_x,k_y\rangle = \sum_{x,y} e^{i(k_x x + k_y y)}|x,y\rangle$, $k_x,k_y \in [0,2\pi]$ with eigenvalue

$R^+|k_x,k_y\rangle = e^{-ik_x}|k_x,k_y\rangle$, $L^-|k_x,k_y\rangle = e^{ik_x}|k_x,k_y\rangle$, $D^-|k_x,k_y\rangle = e^{-ik_y}|k_x,k_y\rangle$, $U^+|k_x,k_y\rangle = e^{ik_y}|k_x,k_y\rangle$. Therefore in the $|k_x,k_y\rangle$ basis (aka Fourier Picture, Fourier Domain) the evolution operator is given by

$$Q(|k_x,k_y\rangle \otimes |\Phi\rangle) = |k_x,k_y\rangle \otimes \left(e^{-ik_x}P_R + e^{ik_x}P_L + e^{-ik_y}P_U + e^{ik_y}P_D\right)M|\Phi\rangle \equiv |k_x,k_y\rangle \otimes M_{k_x,k_y}|\Phi\rangle$$

where $M_{k_x,k_y} = \left(e^{-ik_x}P_R + e^{ik_x}P_L + e^{-ik_y}P_U + e^{ik_y}P_D\right)M$. We should remark that $M_{k_x,k_y}$ is also a unitary operator.

Let $\{A_n\}$ be a set of unital operators on $H_C$, that is, $\sum_n A_n A_n^* = I$. The decoherence on the coin subspace is defined as follows, before each unitary transformation acting on the coin, a measurement given by the unital operators is performed on the coin, after which a density operator $X$ on $H_C$ is transformed by $X \to X' = \sum_n A_n X A_n^*$. The general density operator of the quantum random walk is given by $\rho = \iint \frac{dk_x}{2\pi}\frac{dk_y}{2\pi} \iint \frac{dk'_x}{2\pi}\frac{dk'_y}{2\pi} \left| k_x, k_y \right\rangle \left\langle k'_x, k'_y \right| \otimes X_{k_x,k'_x;k_y,k'_y}$,

where $X_{k_x,k'_x;k_y,k'_y} \in L(H_C)$, and $L(H_C)$ is a vector space of linear operators on $H_C$. After one step of the evolution introduced by the decoherent coin space, the density operator can be written as $\rho' = \iint \frac{dk_x}{2\pi}\frac{dk_y}{2\pi} \iint \frac{dk'_x}{2\pi}\frac{dk'_y}{2\pi} \left| k_x, k_y \right\rangle \left\langle k'_x, k'_y \right| \otimes \sum_n M_{k_x,k_y} A_n X_{k_x,k'_x;k_y,k'_y} A_n^* M^*_{k'_x,k'_y}$. Suppose the quantum walk starts in $\left|0,0\right\rangle \otimes \left|\Phi_0\right\rangle$, then the density operator in the initial state is given by

$\rho_0 = \iint \frac{dk_x}{2\pi}\frac{dk_y}{2\pi} \iint \frac{dk'_x}{2\pi}\frac{dk'_y}{2\pi} \left| k_x, k_y \right\rangle \left\langle k'_x, k'_y \right| \otimes \left|\Phi_0\right\rangle \left\langle \Phi_0 \right|$. After $t$ steps the state evolves to

$\rho_t = \iint \frac{dk_x}{2\pi}\frac{dk_y}{2\pi} \iint \frac{dk'_x}{2\pi}\frac{dk'_y}{2\pi} \left| k_x, k_y \right\rangle \left\langle k'_x, k'_y \right| \otimes L^t_{k_x,k'_x;k_y,k'_y} \left|\Phi_0\right\rangle \left\langle \Phi_0 \right|$, where

$L_{k_x,k'_x;k_y,k'_y} : L(H_C) \mapsto L(H_C)$ is defined by $L_{k_x,k'_x;k_y,k'_y} C \equiv \sum_n M_{k_x,k_y} A_n C A_n^* M^*_{k'_x,k'_y}$ for every

$C \in L(H_C)$. The probability of being at $(x, y)$ at time $t$, is given by

$$P(x,y,t) = Tr\{(\left| x,y \right\rangle \left\langle x,y \right| \otimes I)\rho_t\}$$

$$= \iint \frac{dk_x}{2\pi}\frac{dk_y}{2\pi} \iint \frac{dk'_x}{2\pi}\frac{dk'_y}{2\pi} \left\langle x,y \middle| k_x, k_y \right\rangle \left\langle k'_x, k'_y \middle| x,y \right\rangle Tr\{L^t_{k_x,k'_x;k_y,k'_y} \left|\Phi_0\right\rangle \left\langle \Phi_0 \right|\}$$

$$= \iint \frac{dk_x}{2\pi}\frac{dk_y}{2\pi} \iint \frac{dk'_x}{2\pi}\frac{dk'_y}{2\pi} e^{ix(k_x - k'_x)} e^{iy(k_y - k'_y)} Tr\{L^t_{k_x,k'_x;k_y,k'_y} \left|\Phi_0\right\rangle \left\langle \Phi_0 \right|\}$$

**III. Limiting distribution of coined quantum walk subject to decoherence**

Let $R(q,t) \equiv \left\langle e^{iq(x+y)} \right\rangle_t = \sum_{x,y} e^{iq(x+y)} P(x,y,t)$ be the characteristic function of $P(x,y,t)$. The purpose of this section is to obtain the limit theorems for the decoherent quantum walk. We should

remark that

$$\sum_{x,y} e^{iq(x+y)} P(x,y,t) = \sum_{x,y} e^{iq(x+y)} \iint \frac{dk_x}{2\pi}\frac{dk_y}{2\pi} \iint \frac{dk'_x}{2\pi}\frac{dk'_y}{2\pi} e^{ix(k_x-k'_x)} e^{iy(k_y-k'_y)} Tr\left\{L^t_{k_x,k'_x;k_y,k'_y} |\Phi_0\rangle\langle\Phi_0|\right\}$$

$$= \iint \frac{dk_x}{2\pi}\frac{dk_y}{2\pi} \iint \frac{dk'_x}{2\pi}\frac{dk'_y}{2\pi} \sum_{x,y} e^{-ix(k'_x-k_x-q)} e^{-iy(k'_y-k_y-q)} Tr\left\{L^t_{k_x,k'_x;k_y,k'_y} |\Phi_0\rangle\langle\Phi_0|\right\}$$

$$= \iint \frac{dk_x}{2\pi}\frac{dk_y}{2\pi} \iint \frac{dk'_x}{2\pi}\frac{dk'_y}{2\pi} 4\pi^2 \delta(k'_x-k_x-q, k'_y-k_y-q) Tr\left\{L^t_{k_x,k'_x;k_y,k'_y} |\Phi_0\rangle\langle\Phi_0|\right\}$$

$$= \frac{1}{4\pi^2} \iint \frac{dk_x}{2\pi}\frac{dk_y}{2\pi} Tr\left\{L^t_{k_x,k_x+q,k_y,k_y+q} |\Phi_0\rangle\langle\Phi_0|\right\}$$

where we have used the following property of the dirac delta function

$$\frac{1}{4\pi^2}\sum_{x,y}(xy)^m e^{-ix(k_x-k'_x)} e^{-iy(k_y-k'_y)} = (-i)^m \delta^m(k_x-k'_x, k_y-k'_y).$$

Let $\hat{O} \in L(H_C)$ be any initial state. The generating function of $\left\langle e^{iq(x+y)} \right\rangle_t$ is given by

$$G(z,q) = \sum_{t=0}^{\infty} z^t \left\langle e^{iq(x+y)} \right\rangle_t$$

$$= \iint \frac{dk_x}{2\pi}\frac{dk_y}{2\pi} \sum_{t=0}^{\infty} z^t Tr\left\{L^t_{k_x,k_x+q,k_y,k_y+q} \hat{O}\right\}$$

$$= \iint \frac{dk_x}{2\pi}\frac{dk_y}{2\pi} Tr\left\{\frac{\hat{O}}{I - zL_{k_x,k_x+q;k_y,k_y+q}}\right\}$$

where $|z|<1$ and $\hat{O} \in L(H_C)$. Note that the generating function is well defined by Lemma 1 below.

We should remark that the proof is similar to Lemma 3.1 in Fan et.al [29], therefore we omit it.

**Lemma 1:** Suppose $M \in \{\textit{space of } 4\times 4 \textit{ unitary matrices with complex entries}\}$, and $\{A_n\}$ is a set of unital operators. Let $\lambda$ be an eigenvalue of $L_{k_x,k_x+q;k_y,k_y+q}$, then $|\lambda| \le 1$.

We should remark from Lemma 1 that $\sum_{t=0}^{\infty} z^t L^t_{k_x,k_x+q;k_y,k_y+q} \hat{O}$ converges inside $|z|<1$, and

$I - zL_{k_x,k_x+q;k_y,k_y+q}$ has no poles inside the disk $|z| < 1$. By Lemma 1 and

$I - zL_{k_x,k_x+q;k_y,k_y+q} \in L(L(H_C))$ we have $\left\langle e^{iq(x+y)} \right\rangle_t = \frac{1}{2\pi i} \oint_{|z|=r<1} \frac{G(z,q)}{z^{t+1}} dz$ for some $0 < r < 1$. Since a basis for $L(H_C)$ is given by

$\left\{ \begin{array}{l} \sigma_o \otimes \sigma_o, \sigma_o \otimes \sigma_x, \sigma_o \otimes \sigma_y, \sigma_o \otimes \sigma_z, \sigma_x \otimes \sigma_y, \sigma_x \otimes \sigma_z, \sigma_y \otimes \sigma_z, \sigma_x \otimes \sigma_o, \\ \sigma_y \otimes \sigma_o, \sigma_z \otimes \sigma_o, \sigma_y \otimes \sigma_x, \sigma_z \otimes \sigma_x, \sigma_z \otimes \sigma_y, \sigma_y \otimes \sigma_y, \sigma_z \otimes \sigma_z, \sigma_x \otimes \sigma_x \end{array} \right\}$, where $\sigma_o$, $\sigma_x$, $\sigma_y$, and $\sigma_z$ are the Pauli matrices. We can write $\hat{O}$ as a linear combination of the basis elements.

In column form let us write $\hat{O} = \begin{pmatrix} r_0 \\ r_1 \\ \vdots \\ r_{15} \end{pmatrix}$. Let $A$ be the matrix associated with $I - zL_{k_x,k_x+q;k_y,k_y+q}$, then,

$\frac{\hat{O}}{I - zL_{k_x,k_x+q;k_y,k_y+q}} = A^{-1} \begin{pmatrix} r_0 \\ r_1 \\ \vdots \\ r_{15} \end{pmatrix} = \frac{1}{\det A} \left( A_{ij} \right)_{1 \le i, j \le 16} \begin{pmatrix} r_0 \\ r_1 \\ \vdots \\ r_{15} \end{pmatrix}$, where $A_{ij}$ is the cofactor of $A$. Note that with the exception of $\sigma_0$, the traces of the other Pauli matrices are zero. Since the elements in the basis are in a tensor product, the decomposition implies the trace of $\sigma_0 \otimes \sigma_0$ is four, whilst the traces of the other matrices is zero. Hence when taking the trace in $\iint \frac{dk_x}{2\pi} \frac{dk_y}{2\pi} Tr \left\{ \frac{\hat{O}}{I - zL_{k_x,k_x+q;k_y,k_y+q}} \right\}$, only the the first row action $h(z,q) = \sum_{i=1}^{16} A_{i1} r_{i-1}$ remains. Therefore we have

$G(z,q) = \frac{1}{4\pi^2} \int\int dk_x dk_y \frac{4h(z,q)}{\det A}$. Let $L = \left( l_{ij}(q) \right)$ be the matrix representation of $L_{k_x,k_x+q;k_y,k_y+q}$ in terms of the basis for $L(H_C)$, then we have the following lemma whose proof is similar to Lemma 3.2 in Fan et.al [29], therefore we omit it. We should remark that in Lemma 2 below we have given the matrix representation for $L_{k_x,k_x+q;k_y,k_y+q}$ in terms of the tensor product of the matrix representation for $L_{k,k+v}$ as given in Lemma 3.2 of the paper of Fan et.al [29]. However to get the matrix in Lemma 2 below in terms of $q$, one can use the following relation between $L_{k,k+v}$ and $L_{k_x,k_x+q;k_y,k_y+q}$

$L_{k_x,k_x+q;k_y,k_y+q} = L_{\frac{k_x+k_y}{2},\frac{k_x+k_y+2q}{2}} \otimes L_{\frac{k_x-k_y}{2},\frac{k_x-k_y}{2}}$. We should remark that the proof of Lemma 3.2 in Fan et.al [30] is incomplete, however to get the remaining entries, $X$, it is necessary only to repeat the argument in their proof for each row and/or column with respect to the basis. We should remark that the entries, $X$, in Lemma 3.2 of Fan et.al [29] are not all the same. We make a change in notation in Lemma 2 of the present paper to reflect this.

**Lemma 2:** Suppose $M \in \{space\ of\ 4\times 4\ unitary\ matrices\ with\ complex\ entries\}$, and $\{A_n\}$ is a set of unital operators, then $L_{k_x,k_x+q;k_y,k_y+q}$ has the following representation.

$$\begin{pmatrix} \cos v & a_{12} & a_{13} & a_{14} \\ 0 & a_{22} & a_{23} & a_{24} \\ 0 & a_{32} & a_{33} & a_{34} \\ i\sin v & a_{42} & a_{43} & a_{44} \end{pmatrix} \otimes \begin{pmatrix} \cos v & a_{12} & a_{13} & a_{14} \\ 0 & a_{22} & a_{23} & a_{24} \\ 0 & a_{32} & a_{33} & a_{34} \\ i\sin v & a_{42} & a_{43} & a_{44} \end{pmatrix}.$$

Let us define the probability mass function on $\frac{Z}{\sqrt{t}}$ by $P_t(x,y,t) = P(\sqrt{t}x,\sqrt{t}y,t)$, $x,y \in \frac{Z}{\sqrt{t}}$.

**CLAIM:** $P_t(x,y)$ converges in distribution to a continuous convex combination of normal distributions.

**Proof of Claim:** Let $M \in \{space\ of\ 4\times 4\ unitary\ matrices\ with\ complex\ entries\}$, the unitarity of $M$ implies $|\det M| = 1$, let $\det M = e^{i\gamma}$ then considering the normalized operator $W = e^{\frac{-i\gamma}{2}} M$, implies $M \in SU_2(C) \otimes SU_2(C)$. By the definition of the superoperator $L_{k_x,k'_x;k_y,k'_y}$ and $P(x,y,t)$, $L_{k_x,k_x+q;k_y,k_y+q}$ are the same for $M$ and $W$. Without loss of generality, let $M \in SU_2(C) \otimes SU_2(C)$. By the generating function for $\left\langle e^{iq(x+y)} \right\rangle_t$ and the Cauchy Integral Formula we have

$\left\langle e^{iq(x+y)} \right\rangle_t = \frac{1}{2\pi i} \oint_{|z|=r<1} \frac{G(z,q)}{z^{t+1}} dz = \frac{1}{4\pi^2} \int\int dk_x dk_y \frac{1}{2\pi i} \oint_{|z|=r<1} \frac{4h(z,q)}{z^{t+1} \det A} dz$. Let $z_i(q)$ for $i = 1,\cdots,8$ be

the eight roots of $\det A = 0$. Then, $\frac{1}{z_i(q)}$ for $i = 1, \cdots, 8$ are the eight eigenvalues of $L_{k_x, k_x+q; k_y, k_y+q}$.

Assume 1 is an eigenvalue of $L_{k_x, k_x+q; k_y, k_y+q}$ with multiplicity 1, and $|\lambda| < 1$ for all other eigenvalues $\lambda$ of $L_{k_x, k_x+q; k_y, k_y+q}$ and make the following ordering $1 = |z_0(0)| < |z_1(0)| \leq \cdots \leq |z_8(0)|$. Let $l(z,q) = \frac{4h(z,q)}{z^{t+1} \det A}$. Since for $i = 1, \cdots, 8$ $z_i(q)$ is continuous in $q, k_x, k_y$, there exist $R > 1$ and a small neighborhood $Q$ of $q = 0$ such that $|z_0(q)| < R < |z_i(q)|$ for any $q \in \overline{Q}$, and $k_x, k_y \in [0, 2\pi]$ for all $i = 1, \cdots, 8$. The Cauchy Residue Theorem implies that

$\frac{1}{2\pi i} \oint_{|z|=r<1} \frac{4h(z,q)}{z^{t+1} \det A} dz = \operatorname{Re} s(l, z=0) + \operatorname{Re} s(l, z = z_0(q))$. If $t \geq t_0$, then $t_0$ is such that $\frac{q}{\sqrt{t}} \in \overline{Q}$, for all $q \in (0, 2\pi)$. Let $g(z,q) = \det A$. Since $A = \{(z, q, k_x, k_y) : |z| = R, q \in \overline{Q}, k_x, k_y \in [0, 2\pi]\}$ is compact, there exist constant $C$ such that $\left| \frac{4h\left(z, \frac{q}{\sqrt{t}}\right)}{g\left(z, \frac{q}{\sqrt{t}}\right)} \right| \leq C$ on a compact subset of $A$, for all $t \geq 0$. So $\left| \lim_{t \to \infty} \frac{1}{2\pi i} \oint_{|z|=R} \frac{1}{z^{t+1}} \frac{4h\left(z, \frac{q}{\sqrt{t}}\right)}{g\left(z, \frac{q}{\sqrt{t}}\right)} dz \right| \leq \lim_{t \to \infty} \frac{C}{R^{t+1}} = 0$, thus

$\lim_{t \to \infty} \operatorname{Re} s(l, z=0) = -\lim_{t \to \infty} \operatorname{Re} s\left(l, z = z_0\left(\frac{q}{\sqrt{t}}\right)\right)$. For fixed $q$,

$\operatorname{Re} s\left(l, z = z_0\left(\frac{q}{\sqrt{t}}\right)\right) = \frac{4h\left(z\left(\frac{q}{\sqrt{t}}\right), \frac{q}{\sqrt{t}}\right)}{z_0\left(\frac{q}{\sqrt{t}}\right)^{t+1} \frac{\partial}{\partial z} g\left(z_0\left(\frac{q}{\sqrt{t}}\right), \frac{q}{\sqrt{t}}\right)}$. Let $t \to \infty$, then,

$\lim_{t\to\infty} \operatorname{Re}s\left(l, z = z_0\left(\frac{q}{\sqrt{t}}\right)\right) = -\frac{2h(1,0)}{\frac{\partial g(1,0)}{\partial z}}$. Let $M$ denote the submatrix of $L = (l_{ij})$ given by

$M = \begin{pmatrix} l_{22} & l_{23} & l_{24} \\ l_{32} & l_{33} & l_{34} \\ l_{42} & l_{43} & l_{44} \end{pmatrix} \otimes \begin{pmatrix} l_{22} & l_{23} & l_{24} \\ l_{32} & l_{33} & l_{34} \\ l_{42} & l_{43} & l_{44} \end{pmatrix}$. In view of Lemma 2, the remark prior to the lemma should be considered in expressing $M$ in terms of $q$. Now let $\begin{pmatrix} 1-z & 0 \\ 0 & I_3 - zM|_{v=0} \end{pmatrix} \otimes \begin{pmatrix} 1-z & 0 \\ 0 & I_3 - zM|_{v=0} \end{pmatrix}$ be the matrix associated with $I - zL_{k_x,k_x;k_y,k_y}$ where $\begin{pmatrix} 1-z & 0 \\ 0 & I_3 - zM|_{v=0} \end{pmatrix}$ is the matrix associated with $I - zL_{k,k}$ in Fan et.al [x], and $M|_{v=0} = \left.\begin{pmatrix} l_{22} & l_{23} & l_{24} \\ l_{32} & l_{33} & l_{34} \\ l_{42} & l_{43} & l_{44} \end{pmatrix}\right|_{v=0}$. Since $M|_{v=0}$ is independent of $v$, the decomposition $L_{k_x,k_x;k_y,k_y} = L_{\frac{k_x+k_y}{2},\frac{k_x+k_y}{2}} \otimes L_{\frac{k_x-k_y}{2},\frac{k_x-k_y}{2}}$ is not necessary to write the matrix $\begin{pmatrix} 1-z & 0 \\ 0 & I_3 - zM|_{v=0} \end{pmatrix} \otimes \begin{pmatrix} 1-z & 0 \\ 0 & I_3 - zM|_{v=0} \end{pmatrix}$ in terms of $q$. We should note that the cofactor $A_{i1} = 0$ for $i = 2,3,\ldots,16$, and $A_{11} = \det\left(I_9 - zM|_{q=0}\right)$. In Fan et.al [29] it is shown that $\frac{1}{z_i(0)}$ for $i = 1,2,3$ are the eigenvalues of $M|_{v=0}$, but $M|_{q=0} = M|_{v=0} \otimes M|_{v=0}$, thus the eigenvalues of $M|_{q=0}$ are

$\frac{1}{z_1^2(0)}$, $\frac{1}{z_2^2(0)}$, $\frac{1}{z_3^2(0)}$, $\frac{1}{z_1(0)z_2(0)}$, $\frac{1}{z_1(0)z_3(0)}$, $\frac{1}{z_2(0)z_3(0)}$, where the last three eigenvalues each have multiplicity 2. Hence,

$$\det\left(I_9 - zM|_{q=0}\right) = \left(1 - \frac{z}{z_1^2(0)}\right)\left(1 - \frac{z}{z_2^2(0)}\right)\left(1 - \frac{z}{z_3^2(0)}\right)\left(1 - \frac{z}{z_1(0)z_2(0)}\right)^2$$
$$\left(1 - \frac{z}{z_1(0)z_3(0)}\right)^2\left(1 - \frac{z}{z_2(0)z_3(0)}\right)^2 = \frac{-1}{(z_1(0)z_2(0)z_3(0))^6}(z - z_1^2(0))(z - z_2^2(0))(z - z_3^2(0))(z - z_1(0)z_2(0))^2$$
$$(z - z_1(0)z_3(0))^2(z - z_2(0)z_3(0))^2$$

However,

$$\frac{\partial g(1,0)}{\partial z}=\frac{1}{\left(z_1(0)z_2(0)z_3(0)\right)^6}(1-z_1^2(0))(1-z_2^2(0))(1-z_3^2(0))\left(1-z_1(0)z_2(0)\right)^2$$
$$\left(1-z_1(0)z_3(0)\right)^2\left(1-z_2(0)z_3(0)\right)^2$$

Hence,

$$\frac{2h(1,0)}{\frac{\partial g(1,0)}{\partial z}}=4r_0\frac{\frac{-1}{\left(z_1(0)z_2(0)z_3(0)\right)^6}(z-z_1^2(0))(z-z_2^2(0))(z-z_3^2(0))\left(z-z_1(0)z_2(0)\right)^2\left(z-z_1(0)z_3(0)\right)^2\left(z-z_2(0)z_3(0)\right)^2}{\frac{1}{\left(z_1(0)z_2(0)z_3(0)\right)^6}(1-z_1^2(0))(1-z_2^2(0))(1-z_3^2(0))\left(1-z_1(0)z_2(0)\right)^2\left(1-z_1(0)z_3(0)\right)^2\left(1-z_2(0)z_3(0)\right)^2}=-4r_0=-Tr(\hat{O})=-1$$

since $\hat{O}$ is a density operator. Since $z_0(q)$ is a root of $\det A$, then $g(z_0(q),q)=0$. So

$0=\frac{\partial g(z_0(q),z)}{\partial g}=\left.\frac{\partial g(z,q)}{\partial q}\right|_{z=z_0(q)}+\left.\frac{\partial g(z,q)}{\partial z}\right|_{z=z_0(q)}z_0'(q)$. When $q=0,\ z_0(0)=1$, so

$0=\left.\frac{\partial g(1,q)}{\partial q}\right|_{q=0}+\left.\frac{\partial g(z,0)}{\partial z}\right|_{z=1}z_0'(0)$. Now consider the matrix $A$ at $z=1$ we see that if $i\neq 1,4,13,16$,

then $l_i(q)=0$. So $g(1,q)=(1-\cos^2 q)A_{1,1}-i\sin q\cos q(A_{4,1}+A_{13,1})+\sin^2 qA_{16,1}$. Since the cofactor

$A_{4,1}=A_{13,1}=A_{16,1}=0$, we see that $\left.\frac{\partial g(1,q)}{\partial q}\right|_{q=0}=0$. Since $z_i(0)\neq 0$ for $i=1,2,3$, it follows that

$\left.\frac{\partial g(z_0,q)}{\partial z}\right|_{z=z_0}\neq 0$, hence since $\left.\frac{\partial g(1,q)}{\partial q}\right|_{q=0}=0$, it follows that $\left.\frac{\partial g(z,0)}{\partial z}\right|_{z=1}z_0'(0)=0$ iff $z'(0)=0$

So, $z'(0)=0$ and $\frac{2h(1,0)}{\frac{\partial g(1,0)}{\partial z}}=-1$. From the Dominated Convergence theorem we can get

$\lim_{t\to\infty}R\left(\frac{q}{\sqrt{t}},t\right)=\frac{1}{4\pi^2}\iint_{[0,2\pi]^2}e^{-z_0''(0)q^2}dk_x dk_y$ for $q\in[0,2\pi]$, where $R(q,t)$ is the characteristic function of the probability distribution. From this and the Cramer-Levy Theorem [30] we see that $P_t(x, y)$ converges to a continuous convex combination of normal distributions.

## IV. Concluding Remarks

In this paper we have shown that $P_t(x, y, t)$ converges to a continuous convex combination of normal distributions, under certain eigenvalue conditions. It is an interesting problem to analyze the spectrum of the superoperator $L_{k_x,k_x;k_y,k_y}$ and obtain the necessary and sufficient conditions for the unitary transformation $M$ to satisfy the eigenvalue conditions. Another problem that would be of interest to experimentalist is the performance of the theoretical distribution (limiting distribution) in experiments. Below we discuss the Hadamard walk on the square lattice using neutral atoms trapped in periodic optical potentials to measure the distribution. Since the internal states of the atom are influenced by decoherence resulting from example uncontrollable phase shifts, imperfections in the manipulation by means of laser pulses as well as fluctuations in the trapping potential during lattice shifts, using neutral atoms trapped in optical lattices is a good scenario. In a sense we are roughly proposing parts of the work of Dur et.al [20] to gain insight on how the theoretical distribution performs in experiments. Let the coin operator be the standard two-dimensional Hadamard operator, and consider a single neutral atom at position $(x_0, y_0) = (0,0)$ and the case where the lattice sites $(0,0)$, $(0,1)$, $(1,0)$, and $(1,1)$, which traps the internal states $|00\rangle$, $|01\rangle$, $|10\rangle$, $|11\rangle$, respectively, of the neutral atom, moves with constant velocities $v_{left} < 0$, $v_{right} > 0$, $v_{up} > 0$, and $v_{down} < 0$, respectively, to the left, right, upward, and downward directions respectively. The initial state of the lattice is such that the minimum of a potential well is located at position $(x_0, y_0) = (0,0)$ at $t_0 = 0$. The lattice movements are used to implement the shift operator, while laser pulses allow one to manipulate the internal state of the atom and thus to select the corresponding trapping potential (and therefore the direction of the movement). Given that the atom is initially prepared in state $\frac{1}{2}\left(|00\rangle + i|01\rangle + i|10\rangle - |11\rangle\right)$ at position $(x_0, y_0) = (0,0)$ the application of the Hadamard operator to the internal state of the atom at times $t$ readily implements the quantum random walk by adapting the set up in Dur et.al [20] to the square lattice. The spatial probability distribution of the atom at time $t$ corresponds to the theoretical distribution. The distribution can be measured using a simple fluorescence measurement with several repetitions of the experiment.